\documentclass{new_tlp}

\usepackage{times}
\usepackage{soul}
\usepackage{url}
\usepackage[hidelinks]{hyperref}
\usepackage[utf8]{inputenc}
\usepackage[small]{caption}
\usepackage{graphicx}
\usepackage{amsmath}
\usepackage{booktabs}
\usepackage{algorithm}
\usepackage{latexsym}
\usepackage{stmaryrd}

\usepackage{mathptmx}
\usepackage{quoting}

\usepackage{thm-restate}
\declaretheorem[numberwithin=section]{theorem}
\declaretheorem[numberwithin=section]{corollary}
\declaretheorem[numberwithin=section]{proposition}
\declaretheorem[numberwithin=section]{definition}
\declaretheorem[numberwithin=section]{example}
\declaretheorem[numbered=yes]{lemma }

\usepackage{times}
\usepackage{soul}
\usepackage{url}
\usepackage{graphicx}
\usepackage{amsmath}
\usepackage{booktabs}
\usepackage{algpseudocode}

\usepackage{color}
\usepackage{stackengine}

\usepackage[T1]{fontenc}
\usepackage{courier}

\usepackage[label font=bf]{subfig}

\usepackage{listings}
\lstdefinestyle{asp-style}{
	language=Prolog,
	frame=lines,
	keywordstyle=\linespread{1.1}\scriptsize\ttfamily,
	basicstyle=\linespread{1.1}\scriptsize\ttfamily,
	breaklines=true,
}
\usepackage[english]{babel}
\usepackage{xspace}
\usepackage{amssymb}
\usepackage{pdfsync}
\usepackage{paralist}
\usepackage{srcltx}
\usepackage[acronym,nonumberlist,nohypertypes={acronym}]{glossaries}

\usepackage{tikz}
\usepackage{subfig}
\usetikzlibrary{external}
\usepackage{pgfplots}
\usepgfplotslibrary{external}
\usetikzlibrary{plotmarks}
\usetikzlibrary{pgfplots.groupplots}
\pgfplotsset{
	filter discard warning=false 
	, legend cell align=left
	, minor grid style={loosely dotted, lightgray}
	, major grid style={loosely dashed, lightgray}
}
\usepackage{relsize}

\usepackage[normalem]{ulem}

\usepackage{xcolor}

\newif\ifrevisionmode
\revisionmodefalse

\ifrevisionmode
    \newcommand{\add}[1]{{\color{blue}#1}}
    \newcommand{\remove}[1] { {\color{red}\sout{#1}} }
    \newcommand{\replace}[2]{{\color{red}\sout{#1}}{\color{blue}#2}}
\else
    \newcommand{\add}[1]{#1}
    \newcommand{\remove}[1]{}
    \newcommand{\replace}[2]{#2}
\fi
\def\oursize{\normalsize}

\newcommand{\todo}[1]{{\textcolor{red}{[{{\bf TODO: }#1]}}}}

\newacronym{dlp}{\textsc{DLP}}{{Disjunctive Logic Programming}}
\newacronym{asp}{\textsc{ASP}}{{Answer Set Programming}}
\newacronym{opt}{\textsc{OPT}}{{Overgrounded Program with Tailoring}}

\def\Heads{\ensuremath{Heads}}

\def\HOM{hom}

\def\derives{\ensuremath{\leftarrow}}

\def\newembedding{tailored embedding\xspace}
\def\Newembedding{Tailored embedding\xspace}

\def\newovergrounds{overgrounded programs with tailoring\xspace}
\def\Newovergrounds{Overgrounded programs with tailoring\xspace}

\def\HA{Heads\xspace}
\def\Facts{Facts}

\newcommand{\grnd}{\ensuremath{grnd}\xspace}
\newcommand{\grndt}{\ensuremath{(grnd(P) \cup F)}}

\newcommand{\X}{{\bf X}}
\newcommand{\x}{{\bf x}}

\def\TG{\ensuremath{T\!G}}

\def\UF{\ensuremath{AF}}
\newcommand{\SF}{\ensuremath{PF}}
\newcommand{\OF}{\ensuremath{OF}}
\newcommand{\NF}{\ensuremath{NF}}

\newcommand{\NR}{\ensuremath{NR}}

\newcommand{\DG}{\ensuremath{DG}}

\newcommand{\veel}{ \ |\ }

\newcommand{\nop}[1]{}

\def\embeds{\vdash}
\def\nembeds{\nvdash}
\def\tailors{\Vdash}
\def\EPS{{{\bf ES}}}
\newcommand{\EP}[1]{\ensuremath{{E}_{#1}}}
\newcommand{\E}{\ensuremath{{E}}}

\def\TEPS{{{\bf TE}}}
\newcommand{\TEP}[1]{\ensuremath{{T}_{#1}}}

\def\pacman{{Pac-Man}\xspace}
\def\mmedia{{Content Caching}\xspace}

\newcommand\quo[1]{``#1''}

\newcommand{\inst}{{\ensuremath Inst}}
\newcommand{\instt}[2]{{\ensuremath{{ Inst}({#1},{#2})}}}
\newcommand{\insttinf}[2]{{\ensuremath{{Inst}^\infty({#1},{#2})}}}
\newcommand{\insttk}[3]{{\ensuremath{{Inst}^{#3}({#1},{#2})}}}
\def\deltainst{\ensuremath{\Delta{{\textsc{{Inst}}}}}\xspace}
\def\deltadesimpl{\ensuremath{{{\textsc{{Desimpl}}}}}\xspace}
\def\incrinst{\ensuremath{\textsc{{IncrInst}}}\xspace}

\def\desimpl{\textsc{Desimpl}\xspace}

\newcommand{\simpl}{{\ensuremath Simpl}}
\newcommand{\simpliccolo}{{\simpl[1,3]}}

\newcommand{\simplinf}[2]{{\ensuremath{{Simpl}^\infty({#1},{#2})}}}

\newcommand{\sysfont}{\textit}
\newcommand{\dlv}{\sysfont{DLV}\xspace}
\newcommand{\idlv}{{\mbox{\small $\cal I$-\dlv}}\xspace}
\newcommand{\itwodlv}{{\mbox{\small ${\cal I}^2$-\dlv}}\xspace}

\newcommand{\system}{\mbox{\itwodlv-isd}\xspace}
\newcommand{\oldsystem}{\mbox{\itwodlv-no-isd}\xspace}

\newcommand\hcancel[2][black]{\setbox0=\hbox{$#2$}%
\rlap{\raisebox{.30\ht0}{\textcolor{#1}{\rule{\wd0}{0.5pt}}}}#2}

\algdef{SE}[DOWHILE]{Do}{doWhile}{\algorithmicdo}[1]{\algorithmicwhile\ #1}%

\algnewcommand\algorithmicupdates{\textbf{Updates:}}
\algnewcommand\Update{\item[\algorithmicupdates]}
\algrenewcommand{\algorithmiccomment}[1]{\hfill$/\!\!/${\em #1}}

\input{verticalline}

\begin{document}

\title[Incremental maintenance of overgrounded logic programs with tailored simplifications]{Incremental maintenance of overgrounded logic programs with tailored simplifications\thanks{We thank the reviewers of this paper, whose constructive comments helped to improve our work. This work has been partially supported by MIUR under project “Declarative Reason-
ing over Streams” (CUP H24I17000080001) – PRIN 2017, by MISE under project “S2BDW”
(F/050389/01-03/X32) – “Horizon2020” PON I\&C2014-20, by Regione Calabria under project
“DLV Large Scale” (CUP J28C17000220006) – POR Calabria 2014-20.}
}

    \author[G. Ianni, F. Pacenza and J. Zangari]{
        Giovambattista Ianni,
        Francesco Pacenza,
        Jessica Zangari
        \\
        Department of Mathematics and Computer Science, University of Calabria, Rende, Italy \\
        \email{{\em lastname}@mat.unical.it} - \url{https://www.mat.unical.it}
    }
\jdate{}
\pubyear{}
\submitted{5 June 2020}
\revised{23 July 2020}
\accepted{31 July 2020}

\maketitle

\begin{abstract}
The repeated execution of reasoning tasks is desirable in many applicative scenarios, such as stream reasoning and event processing.
When using answer set programming in such contexts, one can avoid the iterative generation of ground programs thus achieving a significant payoff in terms of computing time.
However, this may require some additional amount of memory and/or the manual addition of operational directives in the declarative knowledge base at hand.
We introduce a new strategy for generating series of monotonically growing propositional programs.
The proposed {\em overgrounded programs with tailoring} (OPTs) can be updated and reused in combination with consecutive inputs.
With respect to earlier approaches, our {\em tailored simplification} technique reduces the size of instantiated programs. A maintained OPT slowly grows in size from an iteration to another while the update cost decreases, especially in later iterations.
In this paper we formally introduce tailored embeddings, a family of equivalence-preserving ground programs which are at the theoretical basis of OPTs and we describe their properties. We then illustrate an OPT update algorithm and report about our implementation and its performance. This paper is under consideration in Theory and Practice of Logic Programming (TPLP).

\end{abstract}


\section{Introduction}\label{sec:intro}

A wide range of applicative contexts require to perform continuous reasoning over event streams.
In turn, this requires the repeated execution of a reasoning task over the same fixed logic program, but with changing inputs.
Among these applications, many can be categorized within the stream reasoning field~\cite{DBLP:journals/datasci/DellAglioVHB17} and include, e.g., real-time motion tracking~\cite{DBLP:conf/aaai/SuchanBWS18}, and decision making for agents, robots, and artificial players in videogames~\cite{DBLP:journals/arobots/SaribaturPE19,DBLP:conf/ruleml/CalimeriGIPPZ18}.

In the above applications, performance requirements are often very demanding: for instance, an artificial player, deployed in a real-time videogame, is subject to a very fast flow of input events, yet it is allowed a very limited time for each decision. In the GVGAI competition~\cite{DBLP:conf/aaai/LiebanaSTSL16} this limit is just $40$ milliseconds.
However, engines based on the answer set semantics are good candidates for reasoning in
such domains, as they encompass advanced reasoning features, declarativity and a potentially good performance.



Recall that the typical workflow of Answer Set Programming (ASP) systems consists in an {\em instantiation} (or {\em grounding}) phase and a subsequent {\em solving} (or {\em answer sets search}) phase. In the first step, a {\em grounder} module produces an equivalent propositional program $gr(P\cup F)$ from an input non-ground logic program $P$ and a set of facts $F$; in the latter step, a {\em solver} module applies dedicated search techniques on $gr(P\cup F)$ for computing the actual semantics of $P \cup F$ in the form of {\em answer sets}~\cite{DBLP:journals/aim/KaufmannLPS16}.
Repeated executions, called {\em shots} or {\em iterations}, can be conceptually abstracted to the task of finding the set of answer sets $AS(P \cup F_i)$ for a sequence of input fact sets $F_1, \dots, F_n$.

Both the grounding and solving performance is critical when highly paced repeated executions are required. This stimulated a research effort towards the development of incremental reasoning techniques in the answer set programming community. The {\em clingo} system and its earlier prototypes~\cite{DBLP:journals/tplp/GebserKKS19} allow a designer to procedurally control how and which parts of the logic program at hand must be incremented, updated and evaluated among consecutive shots.
%
This approach introduces ample flexibility but requires a non-negligible knowledge of solver-specific internal algorithms. Nevertheless, declarativity and fast-prototyping capabilities are a priority in many development scenarios, such as the previously mentioned videogame industry.
In this typical setting,
designers look for easy and off-the-shelf scripting solutions, and do not have knowledge of declarative logic programming at all.


Another approach to incremental reasoning under the answer set semantics consists in using overgrounding techniques~\cite{CIPPZ19}. In this work the grounding step is incrementally performed by maintaining an {\em overgrounded program} $G_P$, which is made \quo{compatible} with new input facts by monotonically enlarging it from one shot to another.
Overgrounding is attractive
since no operational statements are required to incrementally drive the computation.
The time performance of this technique is promising:
 an overgrounded program, after some update iterations, {\em converges} to a propositional theory general enough to be reused together with possible future inputs, with no further update required. This virtually eliminates grounding activities in later iterations,
 however
 the performance of solvers could decrease because of larger input programs.
One can think at overcoming the limitations of overgrounding approaches by introducing techniques limiting the number of generated rules and reducing their size by applying known simplification methods for ground logic programs~\cite{DBLP:conf/ecai/GebserKNS08,DBLP:conf/birthday/FaberLP12}.

However, nonobvious technical obstacles prevent a straightforward extension of overgrounding techniques in the above direction:
in general indeed, simplification criteria are applied based on specific inputs. Consider, e.g., if one simplifies a ground program by properly removing  atoms which are known to be true in all answer sets at a fixed shot. This, and more sophisticated simplification techniques can however be invalidated in later shots, as, for instance, if a logical assertion is no longer supported by the current input. Thus,
diverse general questions arise. One could wonder which properties a ground program should have in order to be \quo{reusable} with a family ${\cal F}$ of different inputs; also, it remains open whether a ground program can be modified in a way such that ${\cal F}$ can be enlarged with small computational cost, and how.
The contributions of this paper are:

\begin{inparaitem}
\item[$(i)$:\ ]
    We characterize a class of ground programs equivalent to the theoretical instantiation, called {\em \newembedding{s}}. \Newembedding{s} make it cleaner to deal with equivalence properties of simplified programs. {\em \Newovergrounds} (OPTs in the following) are series of \newembedding{s} that keep a monotonic growth approach, yet permitting simplification techniques.
\item[$(ii)$:\ ]
    We propose a new incremental grounding strategy, allowing to seamlessly adapt and reuse a ground program in consecutive evaluation shots.
    In particular, OPTs are generated by alternating {\em desimplification} steps, taking care of restoring previously deleted and reduced rules and {\em incremental} grounding steps, which add and simplify new rules.
    The maintained program becomes more and more general (i.e., the family of \quo{compatible} input facts becomes increasingly larger) while moving from a shot to the next, and the update activity becomes progressively lighter.
\item[$(iii)$:\ ]
     We implemented the above strategy in the \idlv grounder. We report about the experimental activities we conducted, comparing with our previous overgrounding strategy and with other state-of-the-art systems. Results confirm that grounding times blend over iterations in the incremental setting, and that the performance of solvers takes advantage from the reduced size of OPTs with respect to plain overgrounded programs.
\end{inparaitem}

The tailored overgrounding approach has several advantages both of theoretical and practical relevance: tailored embeddings overcome many limits of the previous notion of embedding~\cite{CIPPZ19}, and can be easily generalized to other semantics for logic programming, such as the well-founded semantics; their monotonic growth allows for easily implementing caching policies; if a grounding task must be interrupted, restarts on a new shot are almost straightforward to be implemented, since almost no rollback is required; the proposed framework is transparent to knowledge designers; highly general, non-optimized code can benefit from tailored overgrounding as there is no need to worry about which parts of logic programs might be more grounding-intensive.
The rest of the paper is structured as follows:
after overviewing our approach and briefly presenting preliminary notions, we introduce the notion of tailored embedding and its properties. We then present OPTs and a maintenance algorithm thereof, and we report about our prototype and its experimental evaluation; we eventually discuss related work before drawing final considerations. Most proofs
are given in the appendix.

\section{Overgrounding with tailored simplifications: an overview}
\label{sec:outline}
As mentioned, canonical ASP systems work by first {\em instantiating} a non-ground logic program $P$ over input facts $F$, obtaining a propositional logic program $gr(P\cup F)$, and then computing the corresponding models, i.e., the set of answer sets $AS(gr(P\cup F))$.
Notably, systems build ${gr(P\cup F)}$ as a significantly smaller and refined version of the theoretical instantiation but preserve semantics, i.e., $AS(gr(P\cup F)) = AS(P \cup F)$.
The choice of the instantiation function $gr$ impacts on both computing time and on the size of the instantiation.
The grounding procedure $gr$ usually maintains a set $PT$ of ``possibly true'' atoms, initialized as $PT = F$; then, $PT$ is iteratively incremented and used for generating only ``potentially useful'' propositional rules, up to a fixpoint.
Grounding techniques evolved considerably since the early systems based on the explicit usage of domain predicates, and nowadays strategies for decomposing programs and for rewriting, simplifying and eliminating rules are of great help in controlling the size of the final instantiation. The reader can refer to related literature for an overview on grounding optimization techniques~\cite{DBLP:conf/ecai/GebserKNS08,DBLP:journals/ia/CalimeriFPZ17,cali-perr-zang-TPLP-optimizing}.
We show our approach with a simple example.
Let us consider the
program
$P_0$ consisting of rules:
\begin{center}\oursize
\begin{tabular}{lll}
$r(X,Y) \ \derives \ e(X,Y),\ not\ ab(X).$ && $r(X,Z) \veel s(X,Z) \ \derives \ e(X,Y),\ r(Y,Z).$
\end{tabular}
\end{center}

\noindent and the set of input facts $F_1 = \{ e(c,a),\ e(a,b),\ ab(c) \}$.
If only constants $a$, $b$ and $c$ are available, the theoretical grounding of $P_0 \cup F_1$ consists of all possible substitutions of variables with constants, obtaining 9 and 27 instances for the two rules, respectively. With a smarter grounding function $gr$, one can assume $F_1$ as the initial set of \quo{possibly true} facts, then generate new rules and new possibly true facts by iterating through positive head-body dependencies, obtaining the ground program $G_1$:

\def\os{\hspace{4mm}}

\begin{center}\oursize
\begin{tabular}{lll}
$r_1 : r(a,b) \ \derives \ e(a,b),\ not\ ab(a).$ & \os &
$r_2 : r(c,b)  \veel s(c,b) \ \derives \ e(c,a),\ r(a,b).$ \\
$r_3 : r(c,a) \ \derives \ e(c,a),\ not\ ab(c).$
\end{tabular}
\end{center}

A more \quo{aggressive} grounding strategy could also cut or simplify rules:
literals identified as definitely true can be eliminated and rules that cannot fire can be deleted.
We can remove facts $e(a,b)$, and $e(c,a)$ from bodies of $r_1$ and $r_2$, respectively, and rule $r_3$ entirely, obtaining $\TG_1$, composed of rules $r'_1$ and $r'_2$:

\begin{center}\oursize
\begin{tabular}{lll}
$r'_1 : r(a,b) \ \derives \ \hcancel{e(a,b)},\ not\ ab(a).$ & \os &
$r'_2 : r(c,b)  \veel s(c,b) \ \derives \ \hcancel{e(c,a)},\ r(a,b).$ \\
$r_3 : \hcancel{r(c,a) \ \derives \ e(c,a),\ not\ ab(c).}$
\end{tabular}
\end{center}

Nevertheless, $\TG_1$ can be seen as less re-usable than $G_1$, as it cannot be easily extended to a program which is equivalent to $P_0$ with respect to different input facts.
Indeed, let us assume that, at some point, a subsequent run requires $P_0$ to be grounded over facts $F_2 = \{ e(c,a),\ e(a,d) \}$. Note that, with respect to $F_1$, $F_2$ features the addition of facts $F^+ = \{ e(a,d) \}$ and the deletion of facts $F^- = \{ e(a,b),\ ab(c) \}$. The fact $e(c,a)$ belongs to both $F_1$ and $F_2$, and can be seen as a \quo{persistent} fact.
On the one hand, $G_1$ can be easily  made valid for input $F_2$ by just adding new rules, which take into account $F^+$ as new possibly true facts.
On the other hand, $\TG_1$ can be of interest in that simplifications make it smaller than $G_1$. However, $r'_1$ in $\TG_1$ and the absence of $r_3$ would cause wrong inferences for input facts $F_2$, since $\TG_1$ is constructed on the assumption that $e(a,b)$ and $ab(c)$ are true.

Our proposed technique allows to adapt a simplified ground program $\TG_x$ to a new input $F_{x+1}$ by iterating a {\em desimplification} step and an {\em incremental} step on $\TG_x$. The desimplified version of $\TG_x$ is enriched with new simplified rules added in the incremental step.
When $F_2$ is provided as input, the desimplification step restores $r_3$ and reverts $r'_1$ to $r_1$:
\begin{center}
\oursize
\begin{tabular}{lll}
$r_1 : r(a,b) \ \derives \ {\bf e(a,b)},\ not\ ab(a).$ & \os &
$r'_2 : r(c,b)  \veel s(c,b) \ \derives \ \hcancel{e(c,a)},\ r(a,b).$ \\
$r_3 : {\bf r(c,a) \ \derives \ e(c,a),\ not\ ab(c).}$
\end{tabular}
\end{center}
Moreover, in the incremental step two new rules $r_4$ and $r_5$ are  added depending on the new fact $e(a,d)$:
\begin{center}\oursize
\begin{tabular}{lll}
${\bf r_4 : r(c,d)  \veel s(c,d) \ \derives \ e(c,a),\ r(a,d).}$ &\os &
${\bf r_5 : r(a,d) \ \derives \ e(a,d),\ not\ ab(a).}$
\end{tabular}
\end{center}
Then, $r_4$ can be simplified by removing $e(c,a)$, whereas $e(a,d)$ can be deleted from $r_5$, obtaining:
\begin{center}\oursize
\begin{tabular}{lll}
${\bf r'_4 : r(c,d)  \veel s(c,d) \ \derives \ \hcancel{e(c,a)},\ r(a,d).}$ &\os &
${\bf r'_5 : r(a,d) \ \derives \ \hcancel{e(a,d)},\ not\ ab(a).}$
\end{tabular}
\end{center}
Thus, $\TG_2=\{r_1, r'_2, r_3, r'_4, r'_5\}$, whereas $r'_4$ and $r'_5$ were not formerly present in $\TG_1$.
We have now that $\TG_2$ is equivalent to $P_0$ when evaluated over $F_2$ as input facts, whereas $\TG_2$ with input facts $F_1$ would cause wrong inferences. Indeed, $r'_5$ is simplified according to new facts belonging to $F_2$ but not to $F_1$.
Nevertheless, if $F_3=F_1$ is submitted as input, the desimplification step would generate $\TG_3$ from $\TG_2$ by reverting the rule $r'_5$ to $r_5$.



One might notice that $\TG_3$ is built assuming $F_1 \cup F_2 \cup F_3$ as possibly true facts, and assuming $F_1 \cap F_2 \cap F_3$ as certainly true facts. Intuitively, the last element of the series $\{ \TG_i \}$ is the one embracing the larger family of inputs for which \quo{compatibility} is guaranteed, and requiring lesser update work in later iterations:
in case a fourth shot is requested over input facts $F_4 = \{ e(a,d),$\ $e(c,a),$\ $e(a,b)\}$, the desimplification step will leave $\TG_3$ unaltered and the incremental step will not generate new rules; this happens since possibly true facts and persistent facts are left unchanged. Thus, $\TG_4 = \TG_3$.
We illustrate next how programs like $\TG_1$, $\TG_2$ and $\TG_3$ are related to each other, and which formal requirements are necessary to develop a correct incremental grounding strategy.

\section{Preliminaries}\label{sec:preliminaries}
We assume to deal with finite programs under the answer set semantics.
%
A program $P$ is a set of rules. A rule $r$ has form:
%
$\alpha_1 \veel \alpha_2 \veel \dots \veel \alpha_k$ \derives  $\beta_1,\ \dots,\ \beta_n,$\ $not$ $\beta_{n+1},\ \dots,$\ $not$ $\beta_m$.\ \
%
\noindent where $k, n, m \geqslant 0$.
$\alpha_1, \dots, \alpha_k$ and $\beta_1, \dots, \beta_m$ are called {\em atoms}.
An atom has form $p(\X)$, for $p$ a predicate name and $\X$ a list of variable names and constants.
A literal $l$ has form $a$ or $not\ a$, where $a$ is an atom.
The {\em head} of $r$ is defined as $H(r) = \{\alpha_1,\ \dots,\ \alpha_k\}$; 
the {\em positive body} of $r$ is defined as $B^+(r) = \{\beta_1, \dots, \beta_n\}$, whereas the {\em negative body} is $B^-(r) = $ $\{not$ $\beta_{n+1}, \dots, $ $not$ $\beta_m\}$.
The {\em body} of $r$ is $B(r) = B^+(r) \cup B^-(r)$; if $B(r) = \emptyset$ and $H(r) = \{ a \}$ for a single atom $a$, then $r$ is said a {\em fact}.
As usual, we deal with safe logic programs, i.e., for any non-ground rule $r \in P$, and for any variable $X$ appearing in $r$, there is at least one atom $a \in B^+(r)$ mentioning $X$.

A program (resp. a rule, a literal, an atom) is said to be {\em ground} if it contains no variables.
%
%
The set of all head atoms in a ground program $G$ is denoted by $\HA(G) = \bigcup_{r \in G} H(r)$; its set of facts is $\Facts(G)$.
We assume to deal with a fixed {\em Herbrand Universe} consisting of a finite set of constants $U$ and with programs that can be combined with separate input facts. Given a program $P$ and a set of facts $F$, both $P$ and $F$ will only feature constants appearing in $U$.
A {\em substitution} for a rule $r \in P$ is a mapping from the set of variables of $r$ to the set $U$.
A {\em ground instance} of a rule $r$ is obtained by applying a substitution to $r$.
Given a logic program $P$, the {\em theoretical instantiation (grounding)} $\grnd(P)$ of $P$ is defined as the finite set of all ground instances of rules in $P$.

We assume the reader is familiar with the notions of interpretations and models, and with the usual notation in the literature; in particular, when an interpretation $I$ models a ground element $e$ (i.e., an atom, a body, a head, a rule) this is denoted by $I \models e$.
Given $P$ and a set of facts $F$, a set of ground atoms $A$ is an
answer set of $P \cup F$ whenever $A$
is a minimal model of the so-called FLP reduct $(\grnd(P) \cup F)^A$ of $\grnd(P) \cup F$~\cite{DBLP:conf/jelia/FaberLP04}.
We denote the set of all answer sets of $P\cup F$ as $AS(P \cup F)$.
In the following we recall some known results. We are given a logic program $P$ and set of facts $F$.

\begin{theorem}\label{theo:instInfCorretto}\cite{DBLP:conf/iclp/CalimeriCIL08}\
Given a set of ground rules $S$, we define the operator $\instt{P}{S} = \{ r \in \grnd(P)\ s.t.\ B^+(r) \subseteq \HA(S) \}$.
We also define $\insttk{P}{F}{1}$ = $\instt{P}{\emptyset \cup F}$, $\dots$, $\insttk{P}{F}{k}$ = $\instt{P}{\HA(\insttk{P}{F}{k-1})\cup F}$.
%
%
The sequence $\insttk{P}{F}{k}$,  $k \geq 1$, converges to the least fixed point $\insttk{P}{F}{f} = \insttinf{P}{F}$ for some finite value $f$. Also, $AS(P \cup F )$ $=$ $AS(\insttinf{P}{F}\cup F)$.
\end{theorem}

We herein recall the notion of embedding, i.e., an instantiation of $P$ which contains a subset of rules of $grnd(P) \cup F$ sufficient to preserve the answer set semantics for some sets of input facts.

\begin{definition}\label{def:embedding}\rm
[{\em Embedding}]~\cite{CIPPZ19}\ For a set of ground rules $R \subseteq (\grnd(P)\cup F)$, and a ground rule $r \in (\grnd(P) \cup F)$, we say that:
\begin{compactitem}
\item
    $R$ embeds $r$ {\em by body}, denoted $R\ \embeds_b r$, if $\forall a \in B^+(r)$ $\exists r' \in R$ s.t. $a \in H(r')$;
\item
    $R$ embeds $r$ {\em by head}, denoted $R\ \embeds_h r$, if $r \in R$;
\item
    $R$ embeds $r$, denoted $R\ \embeds r$, if either $R\ \nembeds_b r$ or $R\ \embeds_h r$.
\end{compactitem}
%
\noindent A set of ground rules $\E \subseteq \grnd(P)\cup F$ is an {\em embedding program} for $P \cup F$, if $\forall r \in \grnd(P)\cup F$, $\E \embeds r$.
\end{definition}
\noindent
Note that embeddings mimic traditional model-theoretic notions for a logic program, but in the context of positive dependencies in ground rules. In particular, for an embedding $E$ and a ground rule $r$, defining $E \embeds r$ $\Leftrightarrow$ $E \nembeds_b r \vee E \embeds_h r$ enforces a dependence from $B^+(r)$ to $H(r)$; similarly, for a model $M$, the statement $M \models r$ $\Leftrightarrow$ $M \not\models B(r) \vee M \models H(r)$ enforces an implicative dependence from $B(r)$ to $H(r)$.
\begin{proposition}
\label{prop:Embequivalence}\rm
{\em [Embedding equivalence]}~\cite{CIPPZ19}\
Given an embedding $\E$ for $P \cup F$, $AS(\E) =$ $AS(\grnd(P)\cup F) =$ $AS(\insttinf{P}{F}\cup F) =$ $AS(\bigcap_{E' \in \EPS} E')$, where $\EPS$ is the set of embeddings of $P \cup F$.
\end{proposition}
\begin{example}
\label{ex:embedding}
Let us consider program $P_0$ and the set of facts $F_1$ as mentioned in Section~\ref{sec:outline}. Then $G_1 = \{ r_1, r_2, r_3 \} \cup F_1$ is an embedding for $P_0 \cup F_1$.
According to Proposition~\ref{prop:Embequivalence},
$G_1$ is also the minimal embedding for $P_0 \cup F_1$.
\end{example}


%
%
\section{Tailored embeddings}

We are given a logic program $P$ and a set of input facts $F$. We will now work with possibly simplified versions of rules of $\grnd(P)$.
For a rule $r\in \grnd(P)\cup F$, a \emph{simplified rule} (or {\em simplified version}) $s$ of $r$  is a rule annotated with the
set
$B^*(s)$,
where
$B^{*}(s) =$ $B(r)\setminus B(s)$.
The rule $r$ is denoted as $hom(s)$, i.e., $r$ is the \emph{homologous rule} of $s$ belonging to the theoretical grounding whose
body is obtainable as
$B(hom(s)) = B(s)\cup B^*(s)$.
\remove{Whenever $B^*(s) = \emptyset$, we have that $hom(s) = s$.}
For a set of simplified rules $S$ we define $hom(S) = \{ r \in grnd(P) \cup F \replace{|}{\mid} \exists s \in S \mbox{ and } hom(s) = r \}$.
\add{A rule $q \in grnd(P)\cup F$ is regarded as a simplified rule with $B^*(q) = \emptyset$ and $hom(q) = q$. Similarly, a set $Q \subseteq grnd(P)\cup F$ is regarded as a set of simplified rules with $hom(Q) = Q$}.


Note that sets of simplified rules are not comparable under set inclusion, although one can consider, e.g.\add{,} \replace{$\{ a \derives b.\ ,\ c \derives d. \}$}{the set of rules $a \derives b$ and $c \derives d$} as a \quo{somewhat smaller} subset of the set composed by rules \replace{$\{ a \derives b,c.\ ,\ c \derives d. \}$}{$a \derives b,c$ and $c \derives d$}. We thus appropriately generalize set inclusion and set intersection to sets of simplified rules.
Given two sets of simplified rules $S$ and $R$, we say that $S$ is a {\em simplified subset} of $R$ ($S \sqsubseteq R$) if for each $s \in S$ there is a rule $r \in R$ s.t.
$B(s) \subseteq B(r)$ and $hom(s) = hom(r)$.
The {\em simplified intersection} $R \sqcap Q$ of two set of simplified rules $R$ and $Q$ is\add{:}

{
\[
 R \sqcap Q = \{ t \mid  r \in R,  q \in Q,
         B(t) = B(r) \cap B(q),\add{B^*(t) = B^*(r) \cup B^*(q)},
         hom(t) = hom(r) = hom(q)
         \}\add{.}
\]

\begin{example}
\label{ex:sqsubinter}
Let us consider rules $r_1, \dots, r_5$ and their primed versions as mentioned in Section~\ref{sec:outline}. \add{We} assume that for each $i = 1 \dots 5$, $hom(r'_i) = r_i$. Given $\TG_3 = \{ r_1, r'_2, r_3, r'_4, r_5 \} $ and $\TG_1 = \{ r'_1, r'_2 \}$, we have that $\TG_1 \sqsubseteq \TG_3$. \replace{Let}{For} $T_0 = \{ r_1,r'_2 \}$\replace{:}{,} the intersection $T_0 \sqcap \TG_1$ is instead the set $\{ r'_1, r'_2 \}$.
\end{example}

\begin{definition}\label{def:simpl} \rm

Given two sets of simplified rules $R$ and $Q$, we define $\simpl(R,Q)$ as an operator working on each simplified rule
$r \in R$
according to the following simplification types.

\begin{enumerate}

\item
\label{type1}
$r$ is removed from $R$, if there is a literal $not\ a \in B^-(hom(r))$
s.t. $a\in Facts(Q)$;
\item
\label{type2}
 $r$ is removed from $R$ if there is a \replace{literal $l \in B^+(hom(r))$ and $l\notin Heads(Q)$}
 {atom $a \in B^+(hom(r))$ and $a\notin Heads(Q)$};
\item
\label{type3}
 we move from $B(r)$ to $B^*(r)$ each \replace{literal $l \in B^+(hom(r))$ s.t. $l\in Facts(Q)$}
 {atom $a \in B^+(hom(r))$ s.t. $a\in Facts(Q)$}.
\end{enumerate}
\end{definition}
Intuitively, the types $1$ and $3$ depend on atoms which are assumed to be certainly true in any answer set; and the type $2$ depends on atoms that are assumed to be certainly false in any answer set.
With slight abuse of notation, we define $\simpl(R)$ as $\simpl(R,R)$. A number $k$ of repeated applications of $\simpl$ to the same set $R$ is  denoted as $\simpl^k(R)$. Note that, for $k \geq 1$, $\simpl^{k+1}(R) \sqsubseteq \simpl^k(R)$: we denote the fixed point reached in finitely many steps by the sequence of values $\simpl^k(R)$ as $\simpl^\infty(R)$.
\add{We trivially extend the operators $\embeds$, $\embeds_h$ and $\embeds_b$ for a set of simplified rules on the left-hand side and for simplified rules on the right-hand side.}
\add{As given next,} a {\em \newembedding} is a set of simplified rules which extend\add{s} the notion of embedding by including the possibility of using simplification operations in order to obtain smaller, yet correct, ground programs.

\begin{definition}\label{def:simplembedding}\label{def:tailoredembedding}\rm
[{\em {\Newembedding}}]\
\replace{For}{Given} a set of simplified rules $R$ and a rule $r \in \grnd(P) \cup F$,
 we say that $R$ {\em tailors} $r$ ($R \tailors r$) if \add{at least} one of the following holds:
\begin{enumerate}
\item
\label{case1:tailor}
    $R\ \embeds r$;
\item
\label{case2:tailor}
    there exists a simplified rule $s \in R$ such that $hom(s) = r$, $R \embeds_h s$ and
    $R\ \embeds_h a$ for each atom $a\in (B^+(r)\setminus B^+(s))$;
    %
    %
    %
    %

    %
    %

\item
\label{case3:tailor}
    \remove{$R \nembeds_h r$, and} there is a literal $not\ a \in B^-(r)$
    and $R \embeds a$.
\ifrevisionmode
\item
\remove{
\label{case4:tailor}
    $R \nembeds_h r$, and there is a literal $l \in B^+(r)$ for which $l \not\in Heads(R)$;
}
\fi
\end{enumerate}
%
\noindent A set of simplified rules $\E$ is a {\em \newembedding} for $P \cup F$, if $\forall r \in \grnd(P)\cup F$, $\E \tailors r$.
\end{definition}

 Intuitively, a \add{ground} rule $r$ is tailored according to the new operator \quo{$\tailors$} either if it is embedded by $R$ or, otherwise, there are in $R$ the conditions for applying one of the possible simplification types to $r$.
Note that \remove{with slight abuse of notation,}
$R \embeds b$ is meant as a shortcut for $R \embeds \{ b \derives \emptyset\}$.

\add{Informally speaking, the notion of tailored embedding overcomes the one of embedding: although remarkably simple and useful, the latter notion lacks the fact that there are many other classes of optimized ground programs which are of interest, both theoretically and practically. In other words, embeddings do not properly formalize smaller, yet equivalence-preserving, ground programs produced by actual grounders.
The new conditions describe equivalence-preserving ground programs in which a ground rule can be shortened or deleted at all, provided it is \quo{tailored}. This narrows the gap between the formalization~\cite{CIPPZ19} and real applications.
}

\begin{example}
\label{ex:tailoredembedding}
Let us consider the ground program $\TG_3 = \{r_1, r'_2, r_3, r'_4, r_5 \}$ as shown in Section~\ref{sec:outline}, and the set of facts $F_1$. $T = \TG_3 \cup F_1$ is a tailored embedding since:
{\em (a)} $T$ tailors $r_1, r_3$, $r_5$ and all the facts in $F_1$, since $T$ embeds all such rules; {\em (b)} $T \tailors r_2$ since $r'_2$ is a simplified version of $r_2$ for which $T \embeds_h r'_2$ and $T \embeds e(c,a)$; {\em (c)} similarly, $T \tailors r_4$ since $T \embeds_h r'_4$ and $T \embeds e(c,a)$.
Any other rule $r \in grnd(P_0) \cup F_1$ is trivially tailored since it holds that $T \nembeds_b r$ thus implying $T \embeds r$.
\end{example}

\Newembedding{s} enjoy a number of interesting properties: an embedding is a \newembedding (Proposition~\ref{prop:emb-isa-te}); a \newembedding is equivalent to
$P \cup F$ (Theorem~\ref{theo:newEmbequivalence}); also, a simplified intersection of \newembedding{s} is a \newembedding  (Proposition~\ref{prop:newIntersectionEP}); and, importantly, the intersection of all tailored embeddings \add{represents the least tailored embedding under simplified set inclusion and }corresponds to an iterative, operational construction made using the $\simpl$ and $\inst$ operators (Theorem~\ref{theo:newembeddingsarecorrect} and Corollary~\ref{coro:intersequivalence}).

\begin{proposition}\label{prop:emb-isa-te}
An embedding $E$ for $P \cup F$ is a tailored embedding for $P \cup F$.
\end{proposition}

\add{\begin{proof}
We observe that given an embedding $E$ for $P \cup F$, for each $r \in grnd(P) \cup F$, we have that $E \embeds r$. Then $E \tailors r$ by the case~\ref{case1:tailor} of Definition~\ref{def:tailoredembedding}.
\end{proof}}

\add{
\begin{example}
\label{ex:an_embedding_is_a tailored}
Let us consider again the example of Section~\ref{sec:outline} and in particular, the ground program $G_1 = \{r_1, r_2, r_3 \} \cup F_1$. $G_1$ is an embedding for $P_0 \cup F_1$ as each rule $r \in grnd(P_0) \cup F_1$ is embedded by $G_1$: if $r\in G_1$, $G_1\embeds_h r$, whereas if $r \in \{grnd(P_0) \cup F_1\setminus G_1\}$, $G_1\nembeds_b r$. Also, $G_1$ is a tailored embedding for $P_0 \cup F_1$ because for each rule $r\in grnd(P_0) \cup F_1$ we have that $G_1\embeds r$, since we can apply the case~\ref{case1:tailor} of Definition~\ref{def:tailoredembedding}. Note that for $P_0 \cup F_1$, the ground program $\TG_1=\{r'_1, r'_2\} \cup F_1$ is a tailored embedding but it cannot be an embedding. Indeed, $\TG_3 \nembeds r_3$ since $\TG_3 \embeds_b r_3$ and $\TG_3 \nembeds_h r_3$ and according to Definition~\ref{def:embedding}, $\TG_3$ had to embed all rules in $grnd(P_0)\cup F_1$ to be an embedding.
\end{example}
}

\begin{theorem}\label{theo:newEmbequivalence}\rm [Equivalence].
Given a \newembedding $\E$ for $P \cup F$, then $AS(\grnd(P)\cup F)$ = $AS(\E)$.
\end{theorem}

\begin{proposition}\label{prop:newIntersectionEP} \rm [Intersection].
Given two \newembedding{s} $\EP{1}$ and $\EP{2}$ for $P \cup F$,
$\EP{1} \sqcap \EP{2}$ is a tailored embedding for $P \cup F$.
\end{proposition}



\begin{theorem}\label{theo:newembeddingsarecorrect}\rm
Let $\TEPS$ be the set of \newembedding{s} of ${P\cup F}$ and ${\cal E} = \insttinf{P}{F}\cup F$. Then,\ \
\label{cor:lfpisanembedding}
\[
\simpl^\infty({\cal E}) = \bigsqcap_{\TEP{} \in \TEPS} \TEP{}.
\]
\end{theorem}

\begin{corollary}\label{coro:intersequivalence} \rm
By combining  Th.~\ref{theo:newEmbequivalence}, Pr.~\ref{prop:newIntersectionEP} and Th.~\ref{theo:newembeddingsarecorrect},
we have that:
\[ AS(P \cup F) = AS\big(\bigsqcap_{\TEP{} \in \TEPS} \TEP{}\,\big) = AS(\simpl^\infty(\insttinf{P}{F}\cup F))\add{.} \]
\end{corollary}

\begin{example}
\label{ex:leasttembedding}
For program $P_0$ and facts $F_1$ of Section~\ref{sec:outline}, the least tailored embedding of $P_0 \cup F_1$ under simplified set inclusion is the set $\{ r'_1, r'_2 \} \cup F_1$.
\end{example}

%
%
\begin{nop}{
\paragraph{Construction of $P^{-}$.}


When moving from shot $k$ to the following shot $k+1$ some previously applied simplifications might be invalidated. In particular, facts which are no longer true, i.e., appearing in the set $OF_k = SF_{k} \setminus SF_{k-1}$, imply a chain of simplifications which must be retracted.

The atoms which might be interested in a {\em desimplification} operation can be identified declaratively, by considering an auxiliary logic program $P^{-}$.
$P^{-}$ is a modified version of $P$, in which atoms in the form $p^{-}(\x)$ correspond to facts which are no longer true, while atoms in the form $p^{=}(\x)$ correspond to
facts which persist from one shot to another.
$P^{-}$ is built as follows.

We introduce for each predicate $p$ appearing in $P$ two homologous predicate names $p^{-}$ and $p^{=}$. For each $r \in P$ with only one atom in the head, i.e., in the form
\\
{\scriptsize \centerline{
$\alpha(\X)$ \derives  $\beta_1(\X_1),\ \dots,\ \beta_n(\X_n),$\ $not$ $\beta_{n+1}(\X_{n+1}),\ \dots,$\ $not$ $\beta_m(\X_{m})$}\ \
}
\noindent
we introduce in $P^{-}$, for all $k, 1 \leq k \leq n$, a rule $d_{k}$ in the form
\\
{\scriptsize \centerline{
$\alpha^{-}(\X) $ \derives  $\beta_1^{=}(\X_1),\ \dots,\ \beta_k^{-}(\X_k) \dots,\ \beta_n^{=}(\X_n)$}
}
\noindent For a set of atoms $S$, let $S^{-}$ and $S^{=}$ be the set of atoms constructed by respectively replacing each atom name $p$ with their homologous name $p^{-}$ and $p^{=}$.

Given a set of persistent facts $SF^=$, and a set of no longer true facts $OF^{-}$, then
the atoms in the form $p^{-}(\x)$, appearing in the unique answer set of $P^{-} \cup \SF^{=} \cup \OF^{-}$ will encode the atoms which might have triggered a simplification operation that needs to be undone. Note that in general $p^-(\x)$ might identify more simplifications than the ones needing retraction. This does not affect however the correctness of $G_{k+1}$.

\todo{mancano risultati di correttezza e spiegare meglio la ratio. se c'è spazio mettere anche esempio}
}\end{nop} 
\section{Overgrounding with tailoring}
\label{sec:overgrounding}
We illustrate in this section our technique for maintaining appropriate series of tailored embeddings which we call {\em \newovergrounds} (OPTs).

In the following, the logic program $P$ will be coupled with a sequence of sets of input facts $F_1, \ldots, F_n$. We aim to incrementally compute the sets $AS(P \cup F_1), \ldots, AS(P \cup F_n)$ by reducing the burden of the grounding step at the bare minimum, especially in later iterations.
We update and maintain one element of the series of OPTs $G_1, \dots, G_n$ via the repeated execution of an incremental instantiation function called \incrinst, and taking as arguments the program $P$, a ground program $G$ and a set of input facts $F$.
At iteration
$1$, we initialize the global sets of ground atoms $D = \UF =\SF = \emptyset$, and we let $G_1 = \incrinst(P,\emptyset,F_1)$. For an iteration $i > 1$, we will set $G_i = \incrinst(P,G_{i-1},F_i)$.

The series $G_1, \dots, G_n$ has three useful properties:
{\em (i)} for each $i$, $G_i \cup F_i$ is a tailored embedding for $P \cup F_i$ and thus $AS(P \cup F_i) = AS(G_i \cup F_i)$;
{\em (ii)} for the shot $i+1$, the \incrinst function obtains $G_{i+1}$ from $G_i$ by means of an iterative process, which repeatedly undoes now invalid simplifications in $G_i$ (the {\em desimplification} step) and then computes additional new rules $\Delta G_{i+1}$ (the {\em incremental grounding} step);
{\em (iii)} $G_{i+1}$ extends $G_i$, as all the rules of $G_i$ appear in $G_{i+1}$ possibly in their desimplified version, i.e., $G_i \sqsubseteq G_{i+1}$.
The global set $D$ collects the  rules that were deleted at some iteration and could be restored later on, whereas $\UF$ and $\SF$ keep record of so called {\em accumulated facts} and {\em persistently true facts}, respectively.
After computing $G_i$ for a shot $i$, we will have that ${\UF} = \bigcup_{1 \leq k \leq i}$ $F_k$ and $\SF = \bigcap_{1 \leq k \leq i} F_k$. Intuitively, $Heads(G_i)$ will represent {\em possibly true atoms} built by applying the $\inst$ operator starting from $\UF$ as initial set of possible atoms. An atom outside $Heads(G_i)\cup {\UF}$ is assumed to be {\em certainly false} and might trigger simplifications of type 2.
Similarly, $Facts(G_i)$ will be taken as the set of {\em certainly true} atoms which allow to apply simplifications of types 1 and 3.

\paragraph{Outline of the \incrinst function.} An abstract version of the \incrinst function is given in the next page. Let us assume to be at iteration $i+1$ for $i>1$. The \incrinst function is composed of a \deltadesimpl step and a \deltainst step.
On the one hand, in the \deltadesimpl step the rules in $G_i$ are possibly {\em desimplified} whereas previously deleted rules are possibly restored.
On the other hand, undeleted rules and new facts $\NF =  F_{i+1} \setminus AF$ can trigger the generation of new rules, which are incrementally processed in the \deltainst step. These new rules are simplified and added to $G_{i+1}$.


The set $\NR$ keeps track of rules restored from $D$ and of new rules added in the \deltainst step. Atoms in $\HA(\NR) \cup \NF$ can invalidate simplifications of type $2$ 
as they represent no longer certainly false atoms. The set $\OF$ is instead used to keep track of atoms that are no longer assumed to be certainly true at the current shot; the atoms in $\OF$ can invalidate former simplifications of types $1$ and $3$.
The  iterative process internal to \incrinst continues until no new rules are added and no new derived facts need to be retracted, i.e., when both $\NR$ and $\OF$ do not change anymore.

\begin{figure}[t!]
\begin{small}
\begin{algorithmic}[1]
\Require{Non-ground program $P$, ground program $G$, input facts $F$}
\Ensure{A desimplified and enlarged ground program}
\Update{the set of deleted rules $D$, collection of sets $\UF$ and $\SF$}
\Function{\incrinst}{$P,G,F$}
\State $\DG=G$,
\State $\NR = \emptyset$, $\NF = F \setminus \UF$, $\OF = \SF \setminus F$
\State $\UF = \UF \cup F$, $\SF = \SF \cap F$
\While{$\NR \cup \NF$ or $\OF$ have new additions}

 \State{// \em \desimpl\ step}
    \ForAll{$r \in D$} \Comment{{undo simpl. types 1 and 2}}
    \State{$L_1 = \{ not\ a \in B^-(r)$ s.t. $a \in \OF \}$}
    \State{$L_2 = \{ a \in B^+(r)$ s.t. $a \in \HA(\NR) \cup \NF  \}$}
    \If{$L_1 \cup L_2 \not= \emptyset$ }
        \State{$D = D \setminus \{r\}$}
        \State{$\NR = \NR \cup \{r\}$}
    \EndIf 
\EndFor 
\ForAll{$r \in \DG$}
    \State{$L_3 = \{ a \in B^+(r)$ s.t. $a \in \OF \}$}
    %
    %
    \ForAll{$l \in L_3$} \Comment{{undo simpl. type 3}}
        \If{$B(r) = \emptyset \wedge \|H(r)\| = 1$}
            \State{$\OF = \OF \cup H(r)$} \label{alg:addtofminus}
        \EndIf
        \State{$B(r) = B(r) \cup \{l\}$}
    \EndFor 
\EndFor  
\State{// \em \deltainst step}
\Do
   %
   %
   \ForAll{$r \in P$}
        \ForAll{$g \in getInstances(r,DG,\NR\cup\NF)$}
            \If{$\simpliccolo( \{ g \},\NR\cup F) = \emptyset$}  \Comment{{$g$ is deleted}}
                \State $D = D \cup \{ g \}$
            \Else
                \State $\NR = \NR \cup \simpliccolo( \{ g \},\NR\cup F)$ \label{alg:simpl13}
            \EndIf
        \EndFor
   \EndFor
   \doWhile{there are additions to $\NR$}
\EndWhile
\State $S = \simplinf{\NR}{\DG \cup \NR \cup F},\ D = D \cup hom(NR)\setminus hom(S)$ \label{alg:simplinf}\\
\Return $\DG \cup S$
\EndFunction
\end{algorithmic}
\end{small}
%
%
\end{figure}

\paragraph{Desimplification step.}
The $\deltadesimpl$ step makes an update on a copy $\DG$ of  the current ground program $G$ in which simplifications of types $1$ through type $3$ are undone. Note that the desimplification might trigger new additions to $\OF$ and $\NR$, which in turn can cascade new desimplifications and/or new incremental additions.
We purposely allow redundant desimplifications: an atom $f \in Facts(G)$ might be added to $\OF$ as soon as a rule $r$ with $H(r) = \{ f \}$ is desimplified (line~\ref{alg:addtofminus}). However, although there can be some other rule $r_f$ in $G$ such that $H(r_f) = \{ f \}$ and $B(r_f) = \emptyset$, the restore operation on $r$ does not affect the correctness of $\DG$.
\paragraph{Incremental grounding step.}
In this step we instantiate and simplify each rule $r \in P$ that can be constructed using the new ground atoms available in $\HA(\NR)\cup \NF$ up to a fixpoint.
The $getInstances$ function processes a non-ground rule $r$, the ground program $\DG$ and the set $\NR \cup \NF$. All possible new matches for the input rule $r$ are differentially obtained and simplified. The $getInstances$ function can be implemented by carefully adapting semi-naive evaluation techniques. This can avoid the generation of duplicated rules, thus saving computation time and memory consumption.
\paragraph{Simplifications.} Our algorithm applies simplifications over new rules $\NR$ only and in two separate moments: {\em (i)} as soon as a new rule is generated (line~\ref{alg:simpl13}) and {\em (ii)} at the end of the main cycle (line~\ref{alg:simplinf}). In the latter case, we apply all simplification types. In the former case, the $\simpliccolo$ operator is meant to apply only simplifications of types 1 and 3. These two simplification types can be applied earlier and can prevent the generation of rules that will be nonetheless deleted later. We observe that we simplify only newly added rules appearing in $\NR$, but with respect to the current value of $F$. This will make $G_{i+1}$ not \quo{compatible} with inputs $F_k$, $1 \leq k \leq i$. Nevertheless, if some $F_k$ appears again as input in a later iteration, the correctness of $G_{i+1}$ can be achieved with a further desimplification step. It is worth noting that a more conservative strategy could consider only simplifications depending on $PF$.

\begin{example}
Let us recall again the example given in Section~\ref{sec:outline} and consider program $P_0$, the intermediate program $\TG_1 = \{ r'_1, r'_2 \}$ and the set of input facts $F_1 =\{ e(c,a),\ e(a,b),\ ab(c) \}$.
$\TG_1 \cup F_1$ is a tailored embedding for $P_0 \cup F_1$.
Assume also that at this stage $D = \{ r_3 \}$.
Given a new set of input facts $F_2 = \{ e(c,a),\ e(a,d) \}$, we have that $\incrinst(P,\TG_1,F_2)$ works as follows: $\NF$ is initially set to $\{ e(a,d) \}$,  $\OF = \{e(a,b),\ ab(c)\}$, and $\DG$ is initially set to $TG_1$.
The $\desimpl$ step will produce the updated set $\DG = \{ r_1,r'_2,r_3 \}$ by modifying $r'_1$ in $r_1$, while the rule $r_3$ is undeleted and moved from $D$ to $\NR$.
The \deltainst step generates the new rules $r'_5$ and $r'_4$ and adds them to $\NR$. $r'_5$ is a simplified version of $r_5$ constructed using the new atom $e(a,d)$, while $r'_4$ is a reduced version of $r_4$ built using the new atom $r(a,d)$.
No further desimplifications and changes to $\DG$, $\NR$ and $\OF$ happen in the next $\desimpl$ and $\deltainst$ steps nor in the final simplification. The set $\{ r_1,r'_2,r_3,r'_4,r'_5 \}$ is eventually returned.
\end{example}
\def\thincrinst{Let $G_1 = \incrinst(P,\emptyset,F_1)$. For each $i$ s.t. $1 < i \leq n$, let $G_i = \incrinst(P,G_{i-1},F_i)$. Then for each $i$ s.t. $1 \leq i \leq n$, $AS(G_i \cup F_i) = AS(P \cup F_i)$.}

\begin{theorem}\label{theo:desimpliscorrect}
\thincrinst
\end{theorem}

The proof of the above theorem, shown in appendix, goes along the lines of showing how to enlarge, under simplified set inclusion, a tailored embedding $G_{i-1} \cup F_{i-1}$ for $P \cup F_{i-1}$ to a tailored embedding $G_i \cup F_i$ for $P \cup F_i$.

\section{Implementation and experimental evaluation}
The tailored overgrounding strategy described above has been implemented by extending the \idlv grounder~\cite{DBLP:conf/aiia/CalimeriFPZ16,cali-perr-zang-TPLP-optimizing} to a version called \itwodlv.
The new system allows to preload a non-ground logic program $P$, to iteratively submit input facts $F_i$, and to obtain $AS(P \cup F_i)$.
 During the process, a simplified subset $\TG$ of $grnd(P)$ is kept in memory.
 Whenever new input $F_{i+1}$ is submitted, $\TG$ is updated according to the tailored overgrounding strategy; a filtering stage then pipes relevant rules to the solver of choice.
 Homologous, simplified and deleted rules are kept track of by adding mark-up to a single copy of each rule.
Our evaluation was conducted in order to assess {\em (a)} the size of inputs fed to solvers and {\em (b)} the evolution
of the performance of the combination of grounder and solver, given also the changing instantiation size. Since sources of choice points are left substantially unchanged by simplification activities, we expected good improvements in performance due to faster solving times for deterministic parts of ground programs. We considered two benchmarks taken from two real world settings with different specific features: \pacman~\cite{DBLP:conf/ruleml/CalimeriGIPPZ18} and \mmedia~\cite{DBLP:conf/icc/BeckBDEHS17,DBLP:journals/tplp/EiterOS19}. The two benchmarks  constitute good and generalizable real cases of incremental scenarios: the \pacman game allows to assess effectiveness of overgrounding for continuous reasoning in the context of videogames, while \mmedia is a typical example of decision making over fast-paced event streams.
Experiments have been performed on a NUMA machine equipped with two $2.8$GHz AMD Opteron 6320 CPUs, with 16 cores and 128GB of RAM. The measurements have been performed using WASP version 3.0.0, clasp integrated in clingo version 5.4.0 and Ticker version 1.0.
We used two grounder versions: \system stands for our new grounder featuring the new incremental simplification and desimplification techniques ({\em isd} in the following), while \oldsystem is a new improved implementation of plain overgrounded programs~\cite{CIPPZ19}, in which isd techniques are disabled.

\begin{figure}[t]
	\centering
		\includegraphics[width=0.48\textwidth,keepaspectratio]{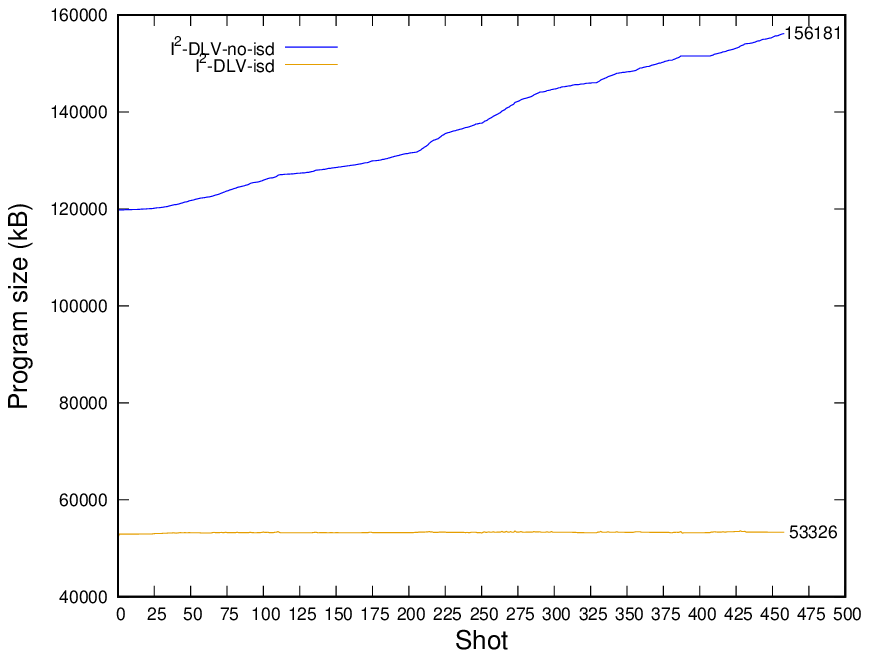} \quad
		\includegraphics[width=0.48\textwidth,keepaspectratio]{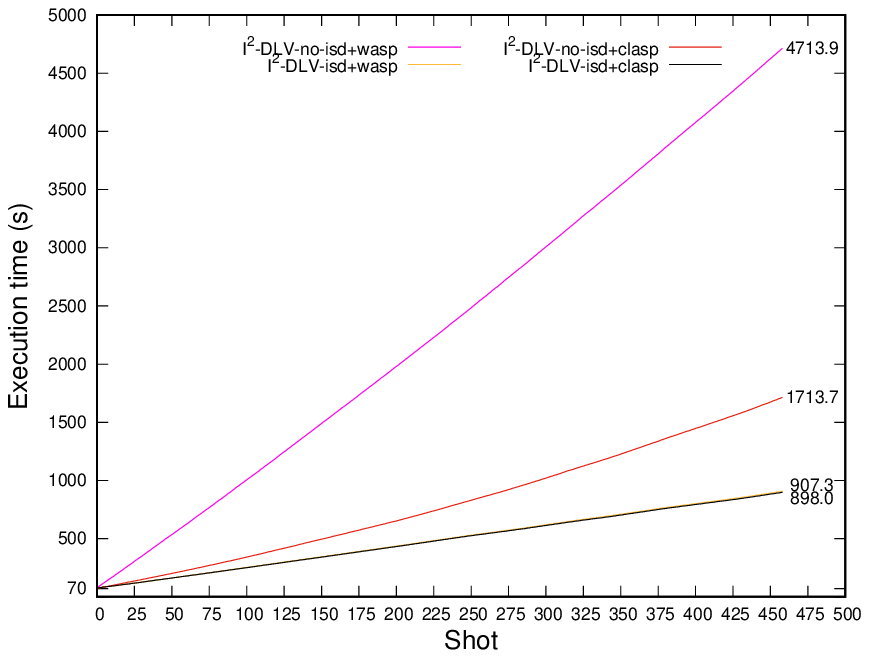}
	\caption{Results of Pac-Man  benchmark. Size on the left side $(a)$ and times on the right side $(b)$.}
	\label{fig:pacman_benchmarks}
\end{figure}

\paragraph{Pac-Man.}
The first experiment has been conducted in the domain of the classic real-time game \pacman. We used the logic program $P_{pac}$  that describes the decision-making process of an artificial player guiding the \pacman in a real implementation~\cite{DBLP:conf/ruleml/CalimeriGIPPZ18}. $P_{pac}$ is repeatedly executed together with different inputs describing the current status of the game map, like e.g., the current position of enemy ghosts, the position of pellets, etc. Several parts of $P_{pac}$ can be considered \quo{grounding-intensive}, like the ones describing the predicate $distance(X_1,Y_1,$ $X_2,Y_2,D)$, where $D$ is computed for all pairs of points $(X_1,Y_1)\times(X_2,Y_2)$, taking care of the shortest path between $(X_1,Y_1)$ and $(X_2,Y_2)$, given the shape of the labyrinth in the game map.
The evaluation has been conducted by logging a series of 
$459$ consecutive sets of input data taken during an actual game;
such inputs were run along with $P_{pac}$ in a controlled environment outside the game engine.
The solving task was performed using WASP~\cite{DBLP:conf/lpnmr/AlvianoDLR15} and clasp~\cite{DBLP:journals/ai/GebserKS12}.
Figure~\ref{fig:pacman_benchmarks}~{\em (a)} compares instantiation sizes for both grounders, while
Figure~\ref{fig:pacman_benchmarks}~{\em (b)} reports about cumulative execution times for the four possible combinations of grounders and solvers. The $X$ axis diagrams data in shot execution order.
Results show that both solvers benefit of the smaller inputs produced by tailored overgroundings, with WASP showing a remarkable improvement. For all the four combinations, the slope of the cumulative time curve reflects an almost constant execution time, with the exception of the first shot measuring around 70 seconds. In this shot, grounding times account for almost all the computation time. A slight progressive worsening in the execution time per shot can be seen especially for combinations involving the old grounder \oldsystem. This is due to the larger program input fed to solvers. For both grounders, we noticed that instantiation times become immediately negligible in later iterations, with \system being around 7\% less performant than \oldsystem because of simplification and desimplification activity.

\paragraph{\mmedia.} In this benchmark, the caching policy of a given video content is controlled using a logic program $P_{cc}$. The caching policy of choice is encoded in the answer sets of $P_{cc} \cup E$ where $E$ encodes a continuous stream of events describing the evolving popularity level of the content at hand. This application has been originally designed in the LARS framework using time window operators in order to quantify over past events~\cite{DBLP:journals/ai/BeckDE18}. We adapted the conversion method specified in the work presenting Ticker~\cite{DBLP:journals/tplp/BeckEB17} to obtain $P_{cc}$ as a plain logic program under answer set semantics, while events were converted to corresponding sets of input facts. These kinds of stream reasoning applications can be fairly challenging, depending on the pace of events and the size of the time window at hand. Our experiments were run in a worst-case scenario in which the caching policy could be decided based on events happening in the last 100 seconds, were the event pace was assumed to be 0.1 seconds. In this setting, a stream reasoning system must be able to deal with a total of $100 \times 10$ different timestamp symbols, and with  proportionally large ground programs. Again, we compared the four combinations of grounders and solvers, and the Ticker system in its two implementations: the Ticker ad-hoc truth maintenance based version (ticker-incr), and the clingo-based one (ticker-asp). Figure~\ref{fig:mmedia_benchmarks}~{\em (a)} shows that both grounders add new rules to their respective overgrounded program up to around shot 1000, which corresponds to the number of time points allowed in the chosen 100 seconds window.
After this threshold, instantiated programs stay constant, with \system generally producing a smaller input. In Figure~\ref{fig:mmedia_benchmarks}~{\em (b)}, the slope of cumulative times shows that ticker-incr has some initial computational cost due to its pre-grounding phase, then performs better in terms of later per-shot times. The four combinations using our grounders have less initial computational cost, while their per-shot times increase slightly in later iterations, with \oldsystem paired with clasp having the best performance, which is quite close to ticker-incr. Ticker-asp does not feature incremental optimization strategies, thus it is not comparable with other solutions.

\begin{figure}[t]
	\centering
		\includegraphics[width=0.48\textwidth,keepaspectratio]{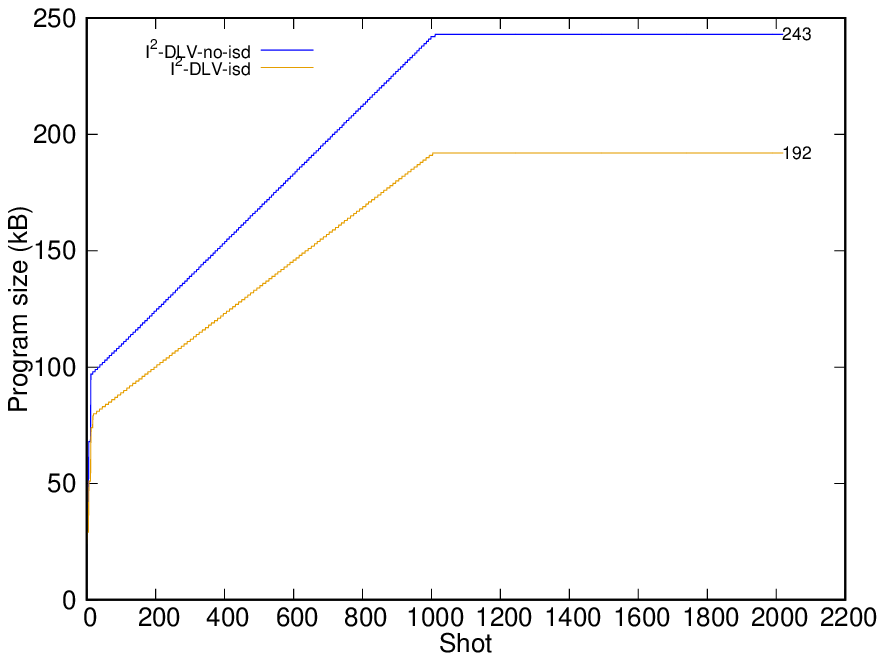} \quad
		\includegraphics[width=0.48\textwidth,keepaspectratio]{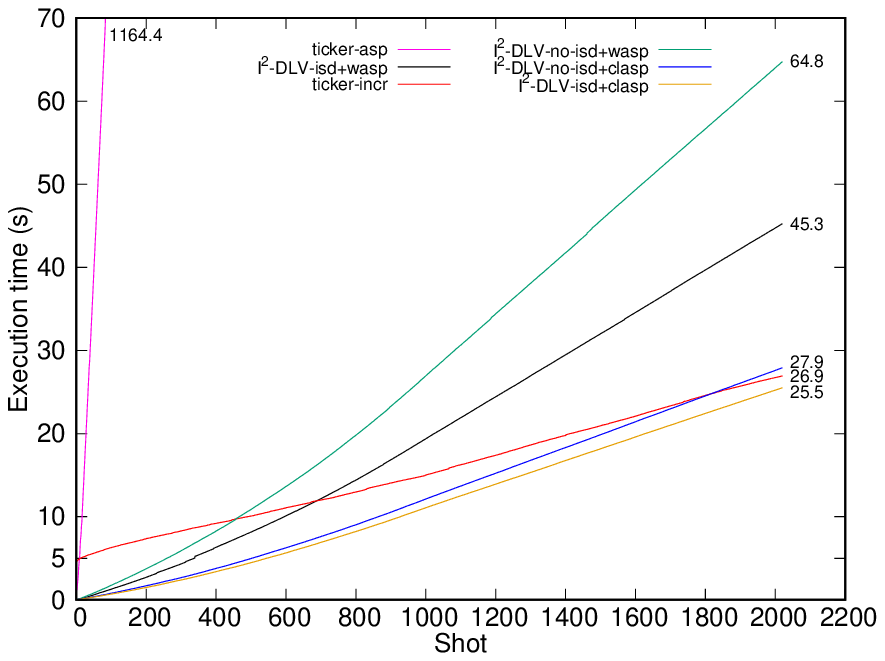}
	\caption{Results of \mmedia benchmark. Size on the left side $(a)$ and times on the right side $(b)$.}
	\label{fig:mmedia_benchmarks}
%
%
\end{figure}

\section{Related work}
\label{sec:relwork}
One of the differentiating ideas of our approach is that OPTs can be \quo{patched} and adapted to new inputs only by adding new information. In other words, OPTs grow monotonically, although an effort has been done to maintain this growth slow.
 A second characteristic of our approach is that we do not include modelling directives for controlling incrementality and multi-shot programs. This choice comes with many advantages (like easier usage and modelling) at the price of the loss of control. In the above respects, our proposal has connections with several lines of research.

It is worth mentioning the early and recent work surveying and proposing incremental update of pool of views~\cite{DBLP:journals/ai/MotikNPH19}.
The aforesaid approaches focus on query answering over stratified Datalog programs and aim to just materialize query answers; our focus is on a generalized setting, where disjunction and unstratified negation are allowed and propositional logic programs are materialized and maintained, for the purpose of computing answer sets.
In contrast with typical delete/rederive techniques,
which require an additional effort to avoid overdeletions and ensure correctness, we purposely allow to perform more desimplifications than necessary. Also, the absence of rederivation activities allows us to keep incremental grounding times low.

In the clingo approach~\cite{DBLP:journals/tplp/GebserKKS19} the notion of incrementality is conceived in a different way:
problems are modelled by thinking in terms of \quo{layers} of modules, for which users specify the grounding and solving sequence thereof.
In the context of overgrounded programs,
modelling can be focussed on a single declarative program. The incremental evaluation over the sequence of input facts is implicit and does not require the user's attention.
Full incremental reasoning, the widest general setting in which a logic program is subject to arbitrary changes and one aims to implicitly maintain answer sets, is to date an almost unexplored topic.
The Ticker system~\cite{DBLP:journals/tplp/BeckEB17} can be seen as a significant effort in this direction, as it implements the LARS stream reasoning formal framework by using back-end incremental truth maintenance techniques under ASP semantics, but is limited to a language fragment with no recursion.
We believe a tighter level of integration between grounders and solvers is necessary in order to achieve full incremental reasoning.
%
Among tightly integrated approaches, it is worth mentioning lazy
grounding~\cite{DBLP:journals/fuin/PaluDPR09,DBLP:journals/tplp/LefevreBSG17,DBLP:conf/aaai/BomansonJW19}. Note that overgrounding is essentially orthogonal to lazy grounding techniques, since these latter essentially aim at blending grounding tasks within the solving step for reducing memory consumption; rather, our focus is on making grounding times negligible on repeated evaluations by explicitly allowing the usage of more memory, while still keeping the two evaluation steps separated.
Finally, it is worth highlighting that \newovergrounds can be seen as an application of relativized hyperequivalent logic programs~\cite{DBLP:journals/tplp/TruszczynskiW09}. A member $G$ of a sequence of OPTs is a logic program which is equivalent to $P$ relative to (part of) a finite set of inputs $F_1, \dots, F_n$. Investigation on the hyperequivalence properties of OPTs, possibly under semantics other than the answer set one, deserves further research.

\section{Conclusions}
\label{sec:conclusions}
Herein we reported about theoretical properties of \newembedding{s} and \newovergrounds; we then presented an algorithm which is at the core of a new incremental grounder. The experiments conducted on our implementation show that smaller instantiations are beneficial for the overall ground \& solve pipeline, and that the grounding effort can be blended over multiple shots, with higher computational impact in earlier iterations. This paves the way to the design of an expressive reasoning system with very short response times, and capable to work over streams of inputs in highly dynamic environments.
The tailored overgrounding strategy is suitable for several extensions. We are currently exploring the possibility of discarding rules when a memory cap is required; also we are investigating towards a cancel/restart technique useful when a reasoning task is aborted.
In both the above scenarios, thanks to the monotonic growth properties of OPTs, partially computed instantiations can be reused on the next shot with almost no rollback burden.
The contexts in which OPTs are beneficial are not restricted to the above, and will be matter of further research.
Detailed experiment results, benchmark encodings, datasets and the binaries of the systems and repeatability information are available at~{\small \url{https://github.com/DeMaCS-UNICAL/I-DLV/wiki/Incremental-IDLV}}. 



%



%
%

\newpage
\appendix

\section{Proofs}
\label{appendix}

\newtheorem{innercustomtheo}{Theorem}
\newenvironment{customtheo}[1]
  {\renewcommand\theinnercustomtheo{#1}\innercustomtheo}
  {\endinnercustomtheo}

\newtheorem{innercustomlemm}{Lemma}
\newenvironment{customlemm}[1]
  {\renewcommand\theinnercustomlemm{#1}\innercustomlemm}
  {\endinnercustomlemm}

\newtheorem{innercustomprop}{Proposition}
\newenvironment{customprop}[1]
  {\renewcommand\theinnercustomprop{#1}\innercustomprop}
  {\endinnercustomprop}

\newtheorem{innercustomcoro}{Corollary}
\newenvironment{customcoro}[1]
  {\renewcommand\theinnercustomcoro{#1}\innercustomcoro}
  {\endinnercustomcoro}

\newtheorem{innercustomdefi}{Definition}
\newenvironment{customdefi}[1]
  {\renewcommand\theinnercustomdefi{#1}\innercustomdefi}
  {\endinnercustomdefi}

Note: in this appendix the numbering of definitions, propositions, lemmas and theorems corresponds to the same statement numbering as in the main paper. Additional statements appearing only in this appendix are labelled with letters. For the sake of readability, statements are repeated together with their full proof, and we recall our assumption that we are given a program $P$ and a set of facts $F$.


\begin{customprop}{A}\label{prop:heads}
For a ground logic program $G$ and $A \in AS(G)$, $A \subseteq \Heads(G)$.
\end{customprop}

\begin{customprop}{B}\label{prop:facts}
For a ground logic program $G$ and $A \in AS(G)$, $\Facts(G) \subseteq A$.
\end{customprop}

The following Proposition re-adapts Theorem~$6.22$~\cite{DBLP:journals/iandc/LeoneRS97}.

\begin{customprop}{C}\label{theo:therearestages}\rm
For a given answer set $A \in AS(P \cup F)$, we can assign to each atom $a \in A$ an integer value $stage(a) =$ $ i$ so that $stage$ encodes a strict well-founded partial order over all atoms in $A$, in such a way that there exists a rule $r \in \grnd(P) \cup F$ structured s.t. 
$a \in H(r)$,
$A \models B(r)$ and
for any atom $b \in B^+(r)$, $stage(b) < stage(a)$.
\end{customprop}

\begin{customprop}{D}\label{theo:therearestagesinembs}\rm
For a given tailored embedding $\E$ for $P \cup F$, let us consider the superset $\Facts(\E)$ of $F$. We can assign to each atom $a \in \Facts(\E)$ an integer value $stage'(a) =$ $ i$ so that $stage'$ represents a strict well-founded partial order over all atoms in $\Facts(\E)$, in such a way that $\HOM(a)$ is structured as follows:
$\{a \} = H(\HOM(a))$,
$\forall b \in B^+(\HOM(a)), b \in \Facts(\E)$ and $stage'(a) < stage'(b)$.
\end{customprop}

\begin{customlemm}{E}\label{lem:facts}\rm
For a tailored embedding $\E$ of $P \cup F$ and an answer set $A \in AS(P \cup F)$, $Facts(\E) \subseteq A$.
\end{customlemm}

\begin{proof}
The proof is given by induction on the function $stage'$ applied to $\Facts(\E)$ as given by Proposition~\ref{theo:therearestagesinembs}.
W.l.o.g. we assign $stage'(a) = 1$ to each atom $a \in \Facts(\E) \cap \Facts(P \cup F)$.
These atoms clearly belong to $A$.
We assume then that for each $a \in \Facts(\E)$ with $stage'(a) < j$ we know that $a \in A$, and show that this implies that for all $a \in \Facts(\E)$ for which $stage'(a) = j$, $a \in A$ as well.
By Proposition~\ref{theo:therearestagesinembs} and the inductive hypothesis, we have that $\HOM(a)$ is such that each $b \in B^+(\HOM(a))$ belongs to $A$, and thus $A \models B^+(\HOM(a))$. Finally, the Lemma is proven by observing that $B^-(\HOM(a)) = \emptyset$.
\end{proof}

\begin{customlemm}{F}\label{lem:stageTorule}\rm
Given a tailored embedding $\E$ of $P \cup F$ and an answer set $A \in AS(P \cup F)$. Then, for each $a \in A$ there exists a rule $r_a \in \E$ s.t. $\HOM(r_a) \in (\grnd(P) \cup F)$; thus, $A \subseteq \HA(\E)$.
\end{customlemm}

\begin{proof}
By Proposition~\ref{theo:therearestages}, each $a \in A$ is associated to an integer value $stage(a)$ and there exists a rule $r_a \in grnd(P) \cup F$, with $a \in H(r_a)$. Note that $r_a \in (grnd(P) \cup F)^A$ since $A \models B(r)$.
We now show that $r_a \in \HOM(\E)$\xspace by induction on the {\em stage} associated to $a \in A$.
W.l.o.g. we can assign $stage(a)=1$, whenever $r_a$ is such that $H(r_a) = \{a\}$, $B^+(r) = \emptyset$ and for all $b$ s.t. $not\ b \in B^-(r)$ we have that $b \notin A$.
When $stage(a)=1$, since $\E$ is a tailored embedding for $P \cup F$, it is easy to check that $\E \embeds_b r_a$, and thus $r_a \in \E$.

Now, (inductive hypothesis) assume that for $stage(a) < j$, $r_a \in \HOM(\E)$.
We show that for $stage(a)= j$, $r_a \in \HOM(\E)$.
Given the above, $r_a$ is such that for each $b \in B^+(r_a)$, $stage(b)< j$, and hence there exists a rule $r_b \in \E$ with $b \in H(r_b)$.
Hence $\E \embeds_b r_a$. Since $\E$ is a tailored embedding for $P \cup F$, and thus $\E \tailors r_a$, we have that at least one of cases in Definition~\ref{def:tailoredembedding} apply. In particular:
\begin{itemize}
\item
If the case~\ref{case1:tailor} applies, $\E \embeds_b r_a$ implies $r_a \in \E$;
\item
If the case~\ref{case2:tailor} applies, there is clearly a rule $r'_a \in \E$ for which $hom(r'_a) = hom(r_a)$;
\item
If the case~\ref{case3:tailor} applies, it must be that for some $not\ b \in B^-(r_a)$,
$b \in Facts(\E)$. But on the other hand $A \models B(r)$ and thus $b \notin A$.
However, by Lemma~\ref{lem:facts}, $b \in A$, which leads to a contradiction.
\ifrevisionmode
\item
\remove{Case~\ref{case4:tailor} cannot apply by induction hypothesis.}
\fi
\end{itemize}
We conclude that either the case~\ref{case1:tailor} or the case~\ref{case2:tailor}, i.e.\add{,} $a \in \Heads(\E)$.
\end{proof}


\begin{customprop}{4.1}
An embedding $E$ for $P \cup F$ is a tailored embedding for $P \cup F$.
\end{customprop}

\begin{proof}
\replace{It is sufficient to observe that an embedding $E$ for $P \cup F$ tailors all the rules of $P \cup F$ by the case~\ref{case1:tailor} of Definition 4.}{The proof is given in the main text.}
\end{proof}

\begin{customtheo}{4.1}
\rm [Equivalence].
Given a \newembedding program $\E$ for $P \cup F$, then $AS(\grnd(P)\cup F)$ = $AS(\E)$.
\end{customtheo}
\begin{proof}
We show that \replace{$A$ is a minimal model of $\E^A$ and $(\grnd(P)\cup F)^A$, both when $AS(\grnd(P)\cup F)$ and when $A \in AS(\E)$}{a given set of atoms $A$ is in $AS(\grnd(P) \cup F)$ iff $A$ is in $AS(E)$}.
We split the proof in two parts.



\paragraph{$\mathbf{[ AS(\grnd(P) \cup F) \subseteq AS(\E)]}$.} Let $A \in AS(\grnd(P) \cup F)$.
We show that $A$ is a minimal model of $\E^A$.
First we show that $A$ is model for $\E^A$. Indeed, let us assume that there is a simplified rule $r \in \E^A$ such that $A \not\models r$. This can happen
only if $A \models B(r)$ but $A \not\models H(r)$.
However, $A \models hom(r)$, which implies that either:
\begin{itemize}
\item
$A \not\models B(hom(r))$. This implies that $\exists l \in B(hom(r))$ such that
$A \not\models l$. We have an immediate contradiction if $l \in B(r)$. Contradiction arises also if $l \not\in B(r)$: indeed, since $\E$ is a tailored embedding, $l$ does not appear in $B(r)$ only if the case~\ref{case2:tailor} of Definition~\ref{def:tailoredembedding} has been applied, which means that a simplification of type 3
has been applied. By Lemma~\ref{lem:facts}, we have a contradiction, since $\Facts(\E) \subseteq A$ implies that $l$ must appear in $A$.

\item
$A \models B(hom(r))$ and thus $A \models H(hom(r))$.
Note that $A \models H(hom(r))$ implies that \mbox{$A \models H(r)$} since $H(r) = H(hom(r))$.
\end{itemize}

We then show that there is no smaller model for $\E^A$. Let us assume that there exist a set $A'$, $A' \subset A$, which is a model for $\E^A$ and thus $A$ is not a minimal model of $\E^A$. Note that $A$ is a minimal model of $(grnd(P) \cup F)^A$ and thus there must exist $r \in \grndt^A$ for which $A' \not\models r$.

Such a rule can be either such that:
\begin{enumerate}[(a)]
\item
\label{p:a}
There is no $s \in \E$ s.t. $r = hom(s)$;
\item
\label{p:b}
There is $s \in \E$ s.t. $r = hom(s)$ and $s \not\in \E^A$;
\item
\label{p:c}
There exists $s \in \E^A$ s.t. $r = hom(s)$.
\end{enumerate}
We show that $r$ cannot fall in the cases {\em~(\ref{p:a})} and {\em~(\ref{p:b})}, while the case {\em~(\ref{p:c})} implies that $A'$ cannot be a model for $\E^A$.

{\em Case~(\ref{p:a})}. Since $r \in \grndt^A$ it is the case that $A \models H(r)$ and $A \models B(r)$. However, by Lemma~\ref{lem:stageTorule}, we know that $A \subseteq \Heads(\E)$. Also, we know that $\E \tailors r$, but there is no $s \in \E$ for which $r = hom(s)$.
This means that $r$ should be tailored either by the case~\ref{case1:tailor}\replace{,~}{ or }\ref{case3:tailor} \remove{or~\ref{case4:tailor}} of Definition~\ref{def:tailoredembedding}.

If the case~\ref{case1:tailor} applies, then it must be that $\E \nembeds_b r$ or $E \embeds_h r$. On the one hand, Lemma~\ref{lem:stageTorule} forces us to conclude that $\E \embeds_b r$; thus it should be the case that $E \embeds_h r$, which contradicts the assumption that $r$ has no $s \in \E$ for which $r = hom(s)$.
If the case~\ref{case3:tailor} applies, \remove{then $\E \nembeds_h r$ and }there exists $not\ a \in B^-(r)$ s.t. $a \in \Facts(\E)$. But by Lemma~\ref{lem:facts}, $\Facts(\E) \subseteq A$, which contradicts $A \models B(r)$.
\remove{In the same way, if the case~\ref{case4:tailor} applies, then $\E \nembeds_h r$ and there exists $a \in B^+(r)$ s.t. $a \notin \Heads(\E)$. But by Lemma~\ref{lem:stageTorule}, $A \subseteq \Heads(\E)$, which contradicts $A \models B(r)$.}

{\em Case~(\ref{p:b})}.
In this case, there is $s \in \E$ s.t. $r = hom(s)$ and $s \not\in \E^A$;
Again, note that $A \models H(r)$ and $A \models B(r)$, which
in turn implies that $A \models B(s)$ and $A \models H(s)$. Thus this case cannot apply, since it turns out that $s \in \E^A$.

{\em Case~(\ref{p:c})}.
Since the two cases above cannot apply, $r$ must fall in this latter case.
Since $A' \not\models r$, it must be the case that
  $A' \not\models H(r)$ and $A' \models B(r)$. Note that $B(s) \subseteq B(r)$ and $H(s) = H(r)$. Thus,
  $A' \not\models H(s)$ and $A' \models B(s)$, which implies
  $A' \not\models s$. We conclude that $A'$ cannot be a model for $\E^A$.

%
%
%
%
%
%
%

\paragraph{$\mathbf{[AS(\E) \subseteq  AS(\grnd(P) \cup F)]}$.}
Let $A \in AS(\E)$. We first show that $A \models \grndt$.
We split all the rules of $\grndt^A$ in two disjoint sets:
$hom(\E^A)$ and $\grndt \setminus hom(\E^A)$.

For a rule $r \in hom(\E^A)$, let $s$ be such that $r = hom(s)$.
We have that $A \models B(s)$ and $A \models H(s)$. Since $H(r) = H(s)$, this latter implies that $A \models H(r)$.
Let us examine each literal $l \in B^*(s)$, which has been eliminated by the case~\ref{case2:tailor} of Definition~\ref{def:tailoredembedding}. \remove{If $l$ is positive }We have that $l \in \Facts(\E)$ , and thus $A \models l$ by Proposition~\ref{prop:facts}.
We can thus conclude that $A \models B(r)$ and, consequently $A \models r$.

Let us now consider a rule $r \in \grndt \setminus hom(\E^A)$. We show that
$A \models r$.
Let us assume, by contradiction that $A \not\models r$, i.e.\add{,} $A \models B(r)$ but $A \not\models H(r)$. We distinguish two subcases: either $r \in hom(\E)$, or $r \not\in hom(\E)$.

If $r \in hom(\E)$, we let $s$ be such that $r = hom(s)$.
Since $r \notin hom(\E^A)$, we have that $s \notin \E^A$, i.e.\add{,}
$A \not\models B(s)$ which implies $A \not\models B(r)$, which contradicts the assumption that $A \not\models r$. If $r \notin hom(\E)$, we however know that $\E \tailors r$. This can be \replace{for three reasons, i.e.}{either because of } the case~\ref{case1:tailor}\replace{,~}{ or the case~}\ref{case3:tailor}\remove{ or~\ref{case4:tailor},} of Definition~\ref{def:tailoredembedding}.

If $r$ falls in the case~\ref{case1:tailor}, we have that $hom(r) = r$ and either $\E \nembeds_b r$ or $\E \embeds_h r$. Since $r \notin hom(\E)$, it must then be that $\E \nembeds_b r$, i.e.\add{,} there exists at least one $a \in B^+(r)$ s.t. it does not exist a rule $r' \in \E$ for which $\E \embeds_h r'$. Then, $a \notin A$ by proposition~\ref{prop:heads} and thus $A \not\models B(r)$.

If $r$ falls in the case~\ref{case3:tailor}, we have that there exist a literal $not\ a \in B^-(r)$ for which $a \in \Facts(\E)$. Clearly, by proposition~\ref{prop:facts}, $a \in A$, and thus $A \not\models B(r)$.
\remove{If $r$ falls in the case~\ref{case4:tailor}, we have that there exist a literal $a \in B^+(r)$ for which $a \notin \Heads(\E)$. Clearly, by proposition~\ref{prop:heads}, $a \notin A$, and thus $A \not\models B(r)$.}

Thus $A \models \grndt^A$.
We know show that $A$ is a minimal model for $\grndt^A$. Let us consider a set
$A' \subset A$ and assume that $A' \models \grndt^A$. However, we know that $A$ is
a minimal model of $\E^A$ and thus $A' \not\models \E^A$. We can show that this implies that $A' \not\models \grndt^A$. Indeed if $A' \not\models \E^A$, then there exists a rule $r \in \E^A$ for which $A' \not\models r$. This, as we will show implies that $A' \not\models hom(r)$ (note that it can be easily shown that $hom(r)$ belongs to $\grndt^A$).

Indeed, we know that $A' \models B(r)$ and $A' \not\models H(r)$. Also it is the case that $A' \models B(r),B^*(r)$. In fact if we assume, by contradiction, that
$A' \not\models B(r),B^*(r)$ there should exist a literal $l \in B^*(r)$ for which $A' \not\models l$. $l$ cannot be negative since $A \models l$ and $A' \subset A$. If $l$ is positive, the case~\ref{case2:tailor} of Definition~\ref{def:tailoredembedding} tells us that $l \in \Facts(\E)$, i.e.\add{,} $\Facts(\E) \not\subset A'$, which in turn implies that $A'$ cannot be a model for $\grndt^A$.
This concludes the proof.
\end{proof}

\begin{customprop}{4.2}
\rm [Intersection].
Given two \newembedding{s} $\EP{1}$ and $\EP{2}$ for $P \cup F$,
$\EP{1} \sqcap \EP{2}$ is a tailored embedding for $P \cup F$.
\end{customprop}
\begin{proof}
Let $\E = \EP{1} \sqcap \EP{2}$, and let us consider a rule $r \in \grndt$. We show that $\E \tailors r$. Preliminarily, we observe two facts which hold by definition of simplified intersection and by the fact that both $\E_1$ and $\E_2$ are tailored embeddings. We are given
a literal $a$ and one of $\E_1$ or $\E_2$ (w.l.o.g., we choose $\E_1$):
\begin{enumerate}[(a)]
\item
\label{pint:B}
$a \in \Facts(\EP{1})$ implies that $a \in \Facts(\E)$.
\item
\label{pint:A}
$a \notin \Heads(\EP{1})$ implies that $a \notin \Heads(\E)$;
\end{enumerate}

By contradiction, let us assume that $\E \not\tailors r$, and we split the proof in two parts, depending on whether $r \in \HOM(\E)$ or whether $r \notin \HOM(\E)$.

($r \in \HOM(\E)$). This implies that there are rules $s \in \E_1$, $q \in E_2$ and $t \in E$ such that $r = hom(s) = hom(q) = hom(t)$. Note that, for each (positive) literal $l \in B^*(t)$, the case~\ref{case2:tailor} of Definition~\ref{def:tailoredembedding} can be applied i.e.\add{,}  $l \in \Facts(\E_1)$ or $l \in \Facts(\E_2)$ which implies $l \in \Facts(\E)$ (Fact~(\ref{pint:B}) above);

($r \notin \HOM(\E)$). In this case we have that either $r \notin \HOM(\EP{1})$ or $r \notin \HOM(\EP{2})$. W.l.o.g. we assume $r \notin \HOM(\EP{1})$. By Definition~\ref{def:tailoredembedding}, this can be the case if either
\begin{enumerate}
\item
$\EP{1} \nembeds_h r$ because there exists $a \in B^+(r)$ and $a \notin \Heads(\EP{1})$. Note that Fact~(\ref{pint:A}) implies that $a \notin \Heads(\E)$, hence $\E \tailors r$.
\item
$\EP{1} \nembeds_b r$; this implies that $\E \nembeds_b r$ hence $\E \tailors r$;
\item
$\EP{1} \nembeds_h r$ because there exists $not\ a \in B^-(r)$, and $a \in \Facts(\EP{1})$. Note that Fact~(\ref{pint:B}) implies that $a \in \Facts(\E)$, hence $\E \tailors r$.
\end{enumerate}

\end{proof}

\begin{customtheo}{4.2}
\rm
Let $\TEPS$ be the set of \newembedding{s} of ${P\cup F}$; let ${\cal E} = \insttinf{P}{F}\cup F$. Then,\ \
\[
\simpl^\infty({\cal E}) = \bigsqcap_{\TEP{} \in \TEPS} \TEP{}.
\]
\end{customtheo}
\begin{proof}
Let ${\cal T} =\bigsqcap_{\TEP{} \in \TEPS} \TEP{} $. By Proposition~2 
we notice that ${\cal E} = \bigsqcap_{\E \in \EPS} \E$. The single argument operator $\simpl$ is both deflationary and monotone when restricted over the complete lattice $( L ,\sqsubseteq)$, where $L = \{ T \in \TEPS | T \sqsubseteq {\cal E} \}$: thus, the iterative sequence $E^0 = sup_{\sqsubseteq}(L) = {\cal E}$, $E^{i+1} = \simpl(E^{i})$ converges to the least fixpoint
$ inf_{\sqsubseteq} ( \{ \TEP{} \in L | \simpl(\TEP{}) \sqsubseteq \TEP{}  \}) =  {\cal T} = \simpl( {\cal T} )$.
\end{proof}
%
%
%
%

\begin{customtheo}{5.1}
\thincrinst
\end{customtheo}
\begin{proof}
The proof is given by induction on the shot indices.
Let $AS_i=AS(P\cup F_i)$. In the base case $(i=1)$, $AS(G_1 \cup F_1) = AS_1$ since the \desimpl\ step has no effect and the $\deltainst$ step coincides with the typical grounding procedure of~\cite{DBLP:conf/birthday/FaberLP12}.
In the inductive case ($i > 1$), we assume that
$G_i \cup F_i$ is a \newembedding for $P \cup F_i$, and we show that
$G_{i+1}  \cup F_{i+1}$ is a \newembedding for $P \cup F_{i+1}$.
Let $G_{i+1} = \incrinst(P,G_i,F_{i+1})$. At the final iteration of the \incrinst algorithm, we have that $G_{i+1} = DG \cup \simpl^\infty(\NR,\DG \cup \NR \cup F_{i+1})$, where $\DG$ is a desimplified version of $G_i$ and $\NR$ is an additional set of rules both obtained by repeated application of \desimpl\ and \incrinst steps.

Observe that
$DG \cup F_i$ is such that $G_i  \cup F_i \sqsubseteq DG_i \cup F_i$ and is a tailored embedding for $P \cup F_i$; then, let $AG_{i+1} = \insttinf{P}{DG \cup F_{i+1}}$.
$DG \cup AG_{i+1} \cup F_{i+1}$ is a tailored embedding for $P \cup F_{i+1}$; it then follows that  $DG \cup \simpl^\infty(AG_{i+1}) \cup F_{i+1}$ is a tailored embedding for $P \cup F_{i+1}$. Let $CG_{i+1} = \{ s \in AG_{i+1} \mid \nexists r \in DG $ s.t. $ hom(r) = hom(s) \}$. $DG_i \cup CG_{i+1} \cup F_{i+1}$ is a tailored embedding for $P\cup F_{i+1}$. Then we show that $CG_{i+1}\sqsubseteq \NR$. It follows that $DG \cup NR \cup F_{i+1} = G_{i+1} \cup F_{i+1}$ is a tailored embedding for $P \cup F_{i+1}$, and that thus $DG \cup \simpl^\infty(\NR,\DG \cup \NR \cup F_{i+1})$ is a tailored embedding for $P \cup F_{i+1}$, which concludes the proof.
\end{proof}


\end{document}